\documentclass[letter]{jpsj3} 
\usepackage{mathrsfs} 
\usepackage{amsmath,amssymb}
\usepackage[]{bm}
\def\bra#1{\left\langle #1 \right|}
\def\ket#1{\left| #1 \right\rangle}
\def\braket#1#2{\left\langle #1 | #2 \right\rangle}
\def\ketbra#1#2{| #1 \rangle \langle #2 | }
\def\rank{{\rm rank}}
\def\comment#1{}

\def\Tr{{\rm Tr}}
\title{Quantum Entanglement of Tensor Networks with Symmetry Projections}

\author{
Masashi \textsc{Orii}\thanks{E-mail address: orii@aquarius.mp.es.osaka-u.ac.jp}, 
Hiroshi \textsc{Ueda}\thanks{E-mail address: ueda@aquarius.mp.es.osaka-u.ac.jp}, 
and Isao \textsc{Maruyama}\thanks{E-mail address: maru@mp.es.osaka-u.ac.jp} 
}

\inst{Department of Material Engineering Science, Graduate School of Engineering Science, Osaka University, 
Toyonaka, Osaka 560-8531, Japan 
}
\abst{
We investigate the global-symmetry projections applied to the tensor network states from the view point of the entanglement entropy and the mutual information. 
The projections to the translational invariant space and to the total-$S^z$-zero space give logarithmically increasing mutual information with respect to the system size. 
In the anti-ferromagnetic $S=1/2$ Heisenberg chain and lattice, 
the optimized energies become accurate numerically by using variational states of the projected tensor network states, 
because the projections reflecting symmetries of the ground states generate quantum entanglement.
}
\kword{
quantum many-body systems, simulation algorithms, tensor networks, quantum entanglement, matrix product state, 
anti-ferromagnetic spin chain, variational method
}
\begin{document}
\maketitle

The design of variational states is an important issue to approach accurate wave function beyond the mean-field analysis in the quantum many body systems.
The matrix and tensor product states (MPS and TPS)~\cite{Ostlund:PRL75, Rommer:PRB55, Niggemann:ZPB104-EPJB13, Delgado:PRB64} 
are suitable variational states for finitely correlated states in one-dimensional (1D) and two-dimensional (2D) systems, respectively,
as used in the density matrix renormalization group (DMRG)~\cite{DMRG}, and the projected entangled pair state (PEPS)~\cite{Verstraete:PRL96}.
The tensor network states, such as the multiscale entanglement renormalization ansatz (MERA) state~\cite{Vidal:PRL101},
are suitable for infinitely long correlation in 1D critical systems to satisfy the entropic area law~\cite{RevModPhys.82.277} of the entanglement entropy (EE)~\cite{PhysRevA.53.2046}.
In other words, a guiding principle for the construction of variational states is to represent large EE within small degrees of freedom (DoF).

In fact, the DMRG becomes less accurate in the critical system than in gapped 1D systems.
This is because the MPS used in the DMRG cannot generate large EE enough to satisfy the entropic area law.
On the other hand, the entropic area law in the 1D critical system is satisfied by the MERA,
which has the disentanglers as an extension of the tree tensor network (TTN) state~\cite{Shi:PRA74}.
In tensor network states including the MPS and TPS, the tensors (or matrices) having variational parameters are connected
and the connection forms a network.
Especially in the MERA and TTN, the networks are tree-like networks spread in spatial dimensions and one additional dimension,
which is a key ingredient to satisfy the entropic area law. 
This network is human-designed but gives a deep insight on an intrinsic holography of the 1D critical system in the sense of the AdS/CFT correspondence\cite{AX.0905.1317}.
In the practical use, the DMRG is still powerful,\cite{SCIENCE.332.1173} because accuracy of each variational method 
depends not only on the type of network but also on the optimization method and computational resources.

One motivation in our preliminary study~\cite{Orii:unpublished}
was to investigate which network is the best independently of optimization schemes,
where the Hamiltonian was the spin $S=1/2$ Heisenberg chain, and the system size was limited to eight sites in order to compare the exact ground state and to compare many networks by one basis of evaluation, i.e., the DoF.
After the optimization of the MPS, TTN, MERA, and some new network states within a given DoF $D$,
accurate variational states are made by the symmetry projections reflecting the global symmetry of the exact ground state,
such as the total $S^z$ conservation, and the translational symmetry.
Due to a merit of small system-size,
the symmetry projections in our preliminary study can be used for any variational states easily compared with the incorporation of the global symmetry into the tensor networks~\cite{PhysRevA.82.050301, PhysRevB.83.115125, PhysRevB.83.115127}.
In addition, there is no additional cost of $D$ in the use of the projections, even if we use two projections simultaneously.
The EE and variational energy get close to the exact values by using the projections irrespective of the details of the network for fixed $D$. 
Refinement due to symmetry projections is usual for the numerical methods~\cite{PRB.72.224518},
but one question arising from the preliminary study is why the projection of 1D translational symmetry gives larger EE than that of the $S^z$ conservation for all networks.
In addition, the 2D translational symmetry cannot be discussed in $N=8$ site system.

Based on this back ground, in this paper, we study numerically the difference of the translational symmetry between 1D and 2D system with $N=16$ sites as a function of DoF $D$
and discuss analytically the projection operators from the view point of the quantum entanglement and the entropic area law. 
To investigate both 1D and 2D systems, we consider the anti-ferromagnetic $S=1/2$ quantum Heisenberg models on the $N=16$ chain and $N=L^2 = 4^2$ square lattice, 
namely $\mathcal{H}_1 = \sum_{i = 1}^{N} {\bm s}_i \cdot {\bm s}_{i + 1}$ and $\mathcal{H}_2 = \sum_{\langle ij,i'j' \rangle}^{} {\bm s}_{ij} \cdot {\bm s}_{i'j'}$, where ${\bm s}$ is the $S=1/2$ spin operator. 
We impose the periodic boundary condition: ${\bm s}_{N+1} = {\bm s}_{1}$ for 1D, ${\bm s}_{L+1,j} = {\bm s}_{1,j}$, and ${\bm s}_{i, L+1} = {\bm s}_{i,1}$ for 2D.
A motivation of this paper is to confirm that variational states used in 1D chain can be applied to 2D lattice with using the projection of 2D translational symmetry.
In the 1D chain, the symmetry projections are applied to the following networks: the 1D spatially uniform MPS, TTN and MERA. 
To study the 2D translational invariant projection, we use the same 1D networks and the spatially uniform TPS for the 2D square lattice.~\cite{note_index_correspondence} 
In addition, we discuss a mutual information~\cite{NielsenChuang} to find out the reason for larger EE given by the projection of 1D translational symmetry than that of the $S^z$ conservation
and show the mutual information has logarithmic dependence on the system size $N$ for projection operators. 

Before showing our numerical results, we define notations following our preliminary work~\cite{Orii:unpublished}.
Projections are denoted as $\hat{P}_{\rm P}$ for total-$S^z$-zero space,  $\hat{P}_{\rm T}$ for total-$k$-zero space,
 $\hat{P}_{\rm B}$ for bond-inversion symmetric space, and
 $\hat{P}_{\rm S}$ for site-inversion symmetric space,
where $k$ is total wave number.
The translational-invariant projection $\hat{P}_{\rm T}$ depends on the dimensionality, i.e.,
$\hat{P}_{\rm T}=\sum_{i=1}^N \hat{T}_x^i/N$ for 1D,
and
$\hat{P}_{\rm T}=\sum_{i=1}^L\sum_{j=1}^L \hat{T}_x^i\hat{T}_y^j /N$ for 2D,
where $\hat{T}_x$ ($\hat{T}_y$) is one-site shift operator along $x$($y$) axis.
Note that one can consider the case of $k\neq 0$ or total $S^z\neq 0$ if the exact ground state has a different symmetry or if you focus on excited states including states with nonzero magetization.
The networks are denoted as $\Psi^{\rm MPS}$, $\Psi^{\rm TPS}$, $\Psi^{\rm TTN}$, and $\Psi^{\rm MERA}$.
Projected variational states are denoted like $\ket{\Psi^{\rm MERA+P+T}}=\hat{P}_{\rm P+T}\ket{\Psi^{\rm MERA}}=\hat{P}_{\rm P}\hat{P}_{\rm T}\ket{\Psi^{\rm MERA}}$ for example.
We impose here a spatially homogeneity of the tensors in the network.
For example, let us explain this spatially homogeneity for the TPS
defined as 
$\Psi^{\rm TPS}({\bm \sigma})=\sum_{{\bm \alpha}^x,{\bm \alpha}^y} \prod_{ij} W_{ij}({\sigma_{ij},\alpha^x_{ij},\alpha^x_{i+1,j},\alpha^y_{ij},\alpha^y_{ij+1}})
$
for a 2D spin configuration ${\bm \sigma}=\{\sigma_{ij}\}$,
where ${\bm \alpha}^{x,y}=\{\alpha_{ij}^{x,y}\}$ are auxiliary DoF and the summation is taken over all integers $\alpha_{ij}^{x,y}\in [1,\chi]$.
The spatial homogeneity in this study imposes uniform tensor $W_{ij}=W$.
Note here that spatially uniform TPS (MPS) is an eigen vector of $\hat{P}_{\rm T}$ in 2D (1D), i.e., $\hat{P}_{\rm T}\ket{\Psi^{\rm TPS}} = \ket{\Psi^{\rm TPS}}$.
The DoF $D$ of the TPS (MPS) comes from one tensor (matrix) $W$ and is a function of $\chi$, which corresponds to the Schmidt rank.
For $N=16$ sites, 1D TTN and MERA have three layers along the artificial dimension and $D$ is determined by ${\bm \chi}=(\chi_1,\chi_2,\chi_3)$.

The variational energy $E=\bra{\Psi}{\cal H}\ket{\Psi}/\braket{\Psi}{\Psi}$ is a function of variational parameters.
The numerical optimization procedure for the network states is the down hill simplex method\cite{Numerical_Recipes:book,Orii:unpublished}.
\begin{figure}[Htb]
  \includegraphics[clip, width=1\columnwidth]{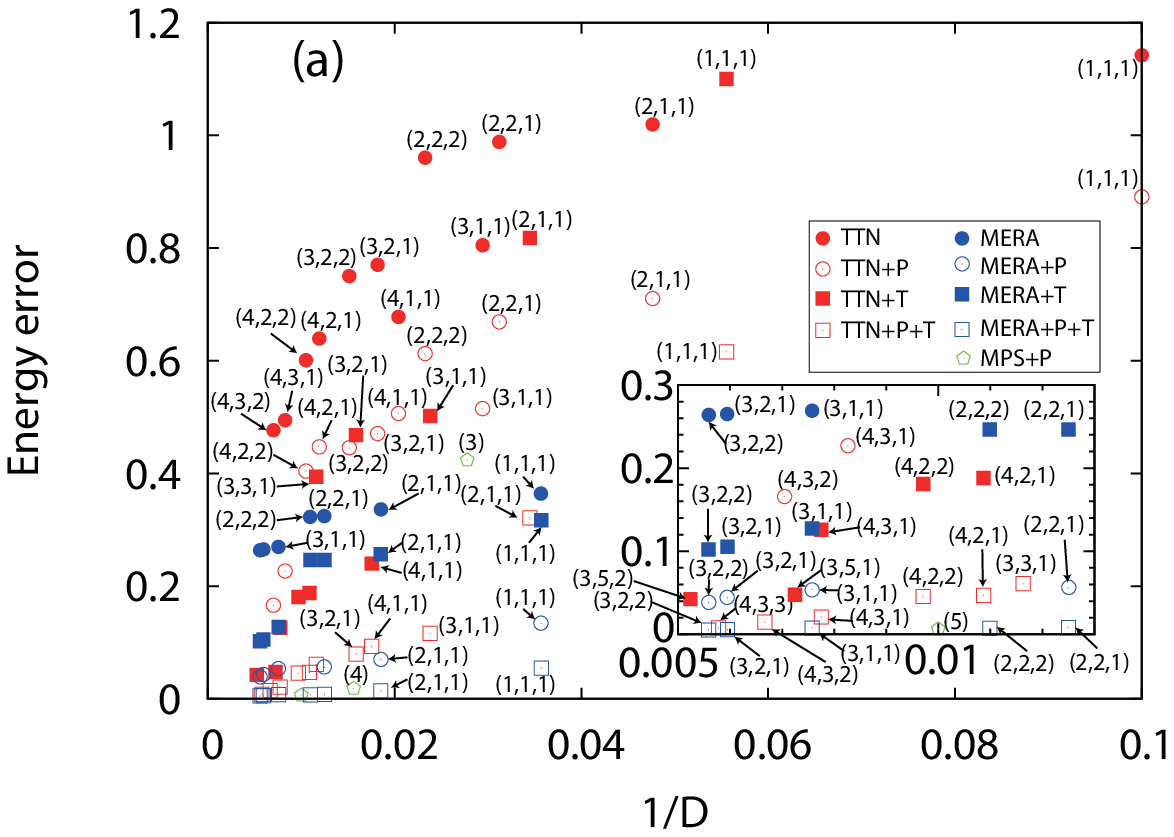}
  \includegraphics[clip, width=1\columnwidth]{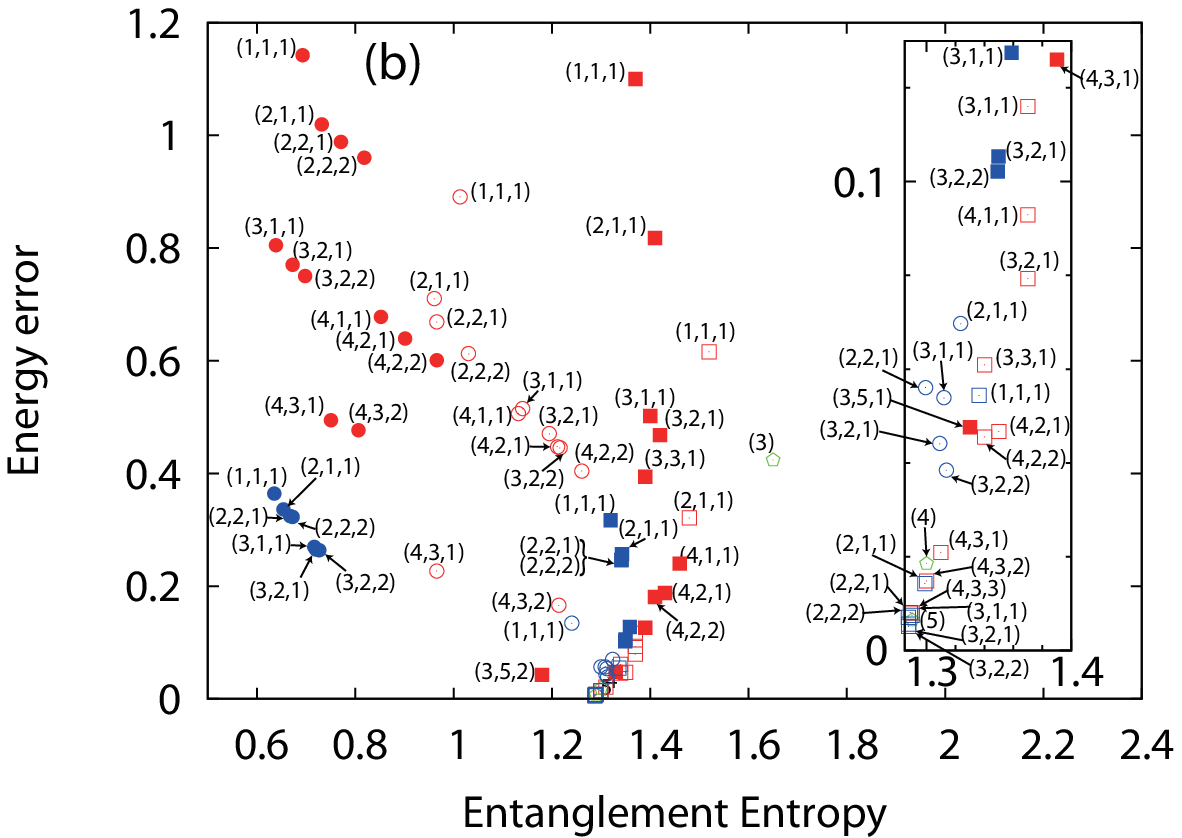}
    \caption{(a) Energy error in the 1D Hamiltonian $\mathcal{H}_1$ as a function of $1/D$, where $D$ is the DoF of the variational wave function,
and (b) energy error as a function of the entanglement entropy. 
Plotted symbols in both figures are common. The system size is 16. 
The parameter $\bm{\chi}$ in each plot is noted by parenthetical numbers. }
    \label{1D}
\end{figure}
Figure \ref{1D} (a) shows the difference between the optimized variational energy and the exact energy in $\mathcal{H}_1$ as a function of $1/D$. 
We again note that the abbreviations +P and +T mean projections into total $S^z=0$ and total $k=0$ spaces respectively.
While the exact ground state is obtained by the variational states in $N=8$ case,
unfortunately we cannot obtain the rigorous energy $E_0=-7.142296$ within $D\leq 179$. 
Therefore, we judge the effectiveness of the network state by smallness of energy error within our parameter space. 
The MERA+T+P of $\bm{\chi}=(3,2,2)$ gives the best variational energy $E=-7.137160$.
However, the best energy in Fig. \ref{1D} (a) is less accurate than the MERA+T+P+B+S of $\bm{\chi}=(3,2,1)$ with additional projections.
Compared with the DMRG numerically, the MERA+T+P (MERA+T+P+B+S) of $\bm{\chi}=(3,2,1)$ roughly corresponds to the DMRG of $\max \chi =15$ ($\max \chi =17$).~\cite{note_DMRG}
Although the results in Fig. \ref{1D} (a) are not accurate as numerical calculations,
it can be concluded that the variational energy is monotonically decreasing by applying each projection operator reflecting the symmetry of the ground state without changing $D$.

The data in Fig. \ref{1D} (a) is replotted as a function of EE with the same symbols in Fig. \ref{1D} (b),
where the EE is bipartite one separating a ring into two chains with the same length and indicates how the bipartite states on two chains are entangled.
Since some networks have position dependent EE while EE of the exact ground state is uniform,
we plot the spatial average of EEs.
While data of the MERA depicted in blue filled circles show a series getting close to the exact value $1.279649$,
the data of the TTN in red filled circles ones strongly depends on the DoF of the bottom layer $\chi_1$.
This is a different point from the $N=8$ case~\cite{Orii:unpublished}.
The same point is that the projection $\hat{P}_{\rm T}$ generates EE more effectively than $\hat{P}_{\rm P}$.

\begin{figure}[Htb]
  \includegraphics[clip, width=1\columnwidth]{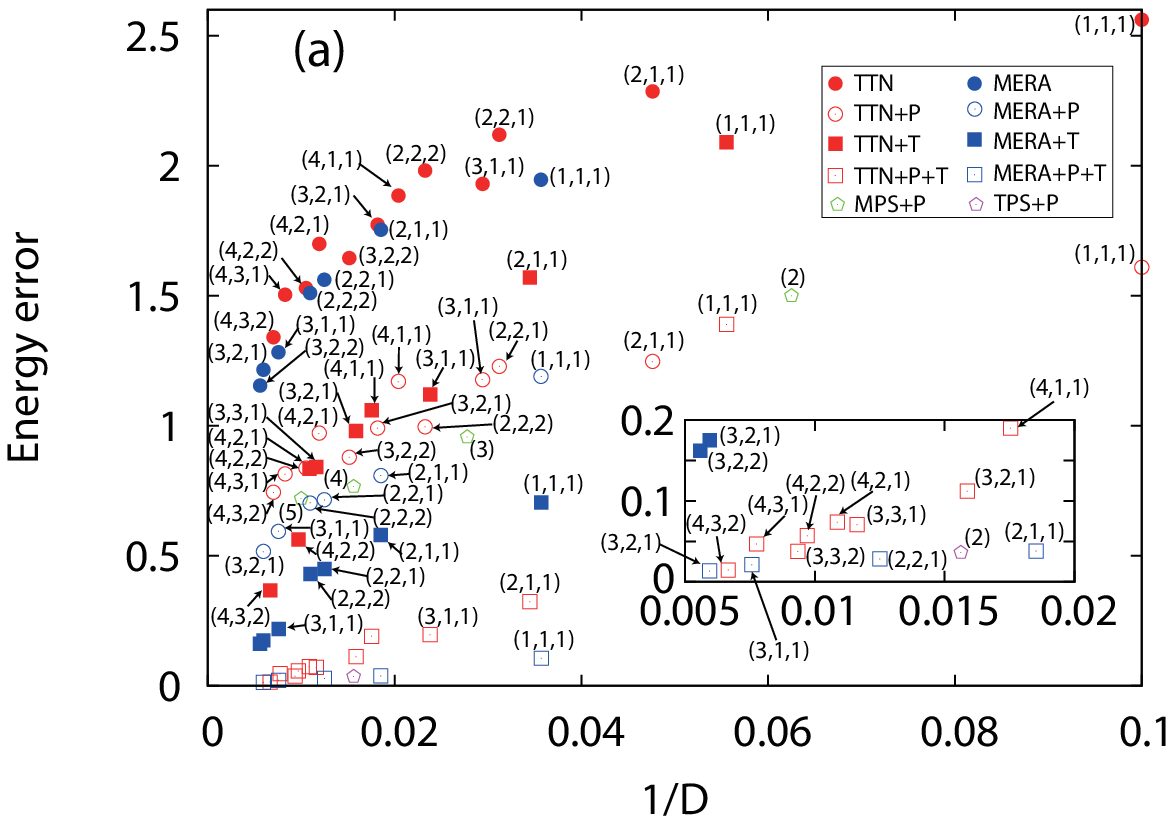}
  \includegraphics[clip, width=1\columnwidth]{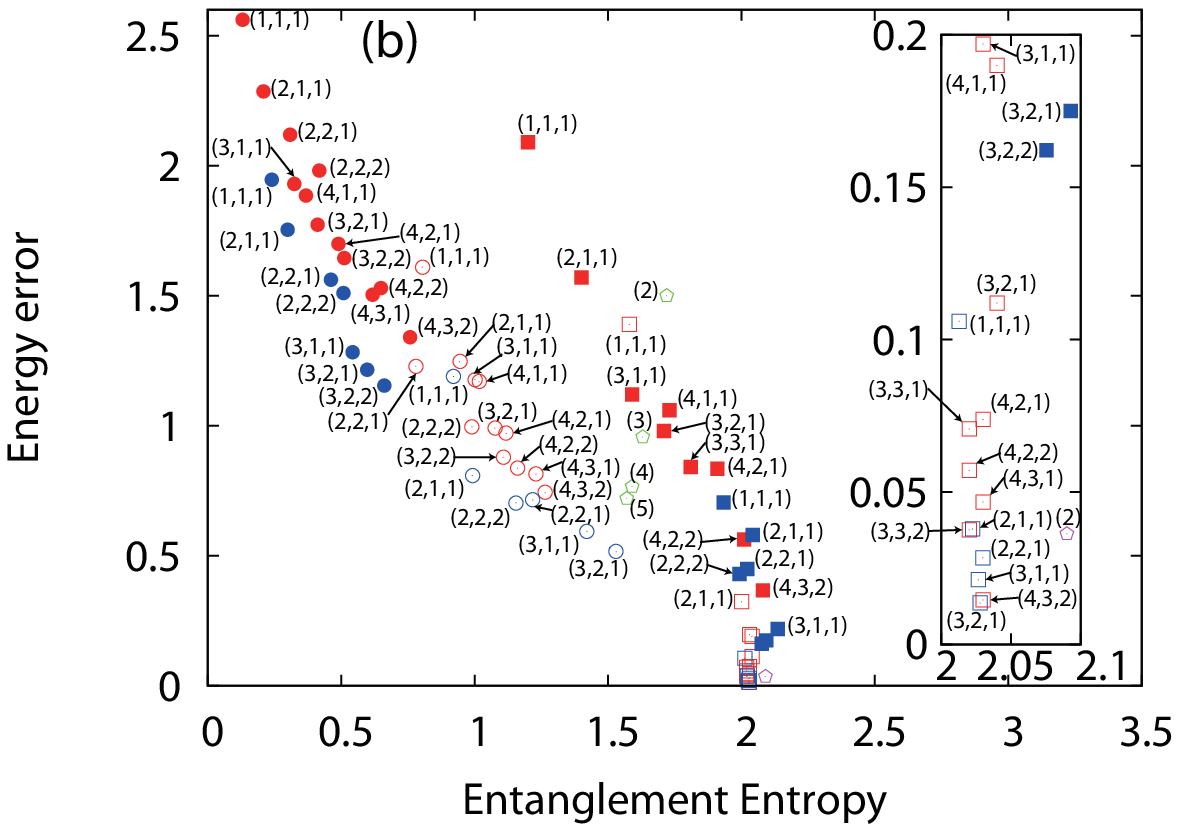}
    \caption{(a) Energy error in the 2D Hamiltonian $\mathcal{H}_2$ as a function of $1/D$, and
(b) energy error as a function of the entanglement entropy. 
Plotted symbols in both figures are the same those in Fig. \ref{1D}.}
    \label{2D}
\end{figure}
As well as the 1D case in Fig. \ref{1D}, we show the numerical results in the 2D Heisenberg square lattice in Fig. \ref{2D},
where the exact energy is $E_0=-11.22848$. 
As shown in Fig. \ref{2D}(a),
the TPS and other tensor networks with $\hat{P}_{\rm T}$ give good variational energy due to reflecting the 2D translational symmetry
compared with the MPS+P, which has 1D translational symmetry.
The importance of 2D translational symmetry is confirmed also in comparison with Fig.~\ref{1D}.
In the 1D case the tensor networks only with $\hat{P}_{\rm T}$ are comparable with those only with $\hat{P}_{\rm P}$ as shown in the inset of Fig.~\ref{1D}
while in the 2D case the networks with $\hat{P}_{\rm T}$ become better than those with $\hat{P}_{\rm P}$ as shown in the inset of Fig.~\ref{2D}.
It should be again noted that the MERA, TTN, and MPS used for Fig. \ref{2D} are the same networks with 1D case~\cite{note_index_correspondence}.
The energy correction is due to $\hat{P}_{\rm T}$ of 2D translational symmetry.
%
In this sense, the dimensionality of systems, which gives big difference on the entropic area law, is absorbed by $\hat{P}_{\rm T}$ at least in $N=16$ sites.
In addition, when we compare the TTN and MERA results, we can know the effect of the disentanglers,
because the networks of the TTN and MERA are almost same but the MERA has disentanglers.
As a result, the disentanglers refine energy error generally both in the 1D and 2D case with $\hat{P}_{\rm P+T}$.
The importance of the translational symmetry is confirmed clearly by the EE shown in Fig. \ref{2D} (b),
where the exact EE is 2.028496.
Since the MPS has 1D translational invariance~\cite{note_index_correspondence}, it shows complete opposite trend.
In general, the EE is recovered by $\hat{P}_{\rm T}$ while the states only with $\hat{P}_{\rm P}$ cannot get close to the exact EE.
From Fig. \ref{2D} (b),
it can be concluded that EE is generated by projected network states 
and the projection of $\hat{P}_{\rm T}$ gives larger EE than that of $\hat{P}_{\rm P}$ in not only 1D but also 2D system. 

%
Here, we discuss why the symmetry projections generate the quantum entanglement.
To simplify following discussion, we consider the 1D case and the EE between two regions A and B and assume the regions have the same number of spins, namely $N_{\rm A} = N_{\rm B} = N/2$. 
To discuss the efficiency of symmetry projections $\hat{P}$ as an entropic generator, we prepare a direct product state (DPS) $\ket{\Psi} = \ket{\psi_{\rm A}} \ket{ \psi'_{\rm B}}$ which has
no EE; $S_{\rm A} = \Tr[ - \rho_{\rm A} \log \rho_{\rm A} ] = 0$ trivially, where $\hat{\rho}_{A} = \Tr_{\rm B} \ketbra{\Psi}{\Psi}$ and $\braket{\Psi}{\Psi}=1$. 
Then, we consider  the projected state $|{\tilde{\Psi}}\rangle=\hat{P}\ket{\Psi}/\sqrt{{\overline N}}$ with the normalization factor ${\overline N}$,
which can have non-zero EE $S_{\rm A}$ because projection generates the entanglement.
The maximum of $S_{\rm A}$ of $\tilde{\Psi}$ is given by 
$\min [ N_{\rm A} \log 2, \log( \rank[\tilde{P}] ) ]$,
where the matrix $\tilde{P}$ is given by the projection operator $\hat{P}$ as
$\bra{{\bm \sigma}'_{\rm A} {\bm \sigma}_{\rm A}}\tilde{P} \ket{{\bm \sigma}'_{\rm B} {\bm \sigma}_{\rm B}}= \bra{{\bm \sigma}'_{\rm A} {\bm \sigma}'_{\rm B}}\hat{P}\ket{{\bm \sigma}_{\rm A} {\bm \sigma}_{\rm B}}$.
Note that this is valid only for DPS state with the projection.

As a trivial case, for the identity operator $\hat{P} = \hat{1}$, we obtain $\rank[\tilde{P}] = 1$ and $S_{A} = 0$.
For $\hat{P} = \hat{P}_{\rm P}$,  we obtain $\max [ S_{\rm A} ] = \log( \rank[\tilde{P}_{\rm P}] ) = \log (N/2 + 1)$. 
This is nothing but an entropic generation by the projection.
In fact, there is the example\cite{note_II}
where $|{\tilde{\Psi}}\rangle$ of $\hat{P}_{\rm P}$ has the maximum EE, $\log(N/2 + 1)$. 
The rank of the 1D translational invariant projection $\hat{P}_{\rm T}$ is $2^{N}$, which indicates $\max[ S_{\rm A}] = (N/2) \log 2$. 
Then, it is concluded that maximum EE generated by the projection from the DPS is larger for $\hat{P}_{\rm T}$ than that for $\hat{P}_{\rm P}$.
The rank of $\tilde{P}$, however, is not a versatile guidepost for evaluation of the entanglement. 
For example, the inversion operator $\hat{T}_{\rm B}$ which substitutes the region A and B
gives $\rank[\tilde{T}_{\rm B} ] = 2^{N/2}$ but $S_{\rm A}$ is always zero because $\tilde{\Psi}$ is the DPS if $\Psi$ is a DPS. 

Therefore, we introduce a mutual information for projection operators $\hat{P}$ using the state $\Psi$ with 
the vector elements defined by 
$\braket{{\bm \sigma}_A{\bm \sigma}_B{\bm \sigma}'_{\rm A\cup B}}{\Psi}=\bra{{\bm \sigma}_A{\bm \sigma}_B}\hat{P}\ket{{\bm \sigma}'_{\rm A\cup B}}/\sqrt{\overline N}
$
where ${\overline N}
$ is the normalization factor
and a region ${\rm A\cup B}$ is the total system.
Then, $\Psi$ is a normalized $2^{2N}$-dimensional vector as in the context of purification\cite{NielsenChuang}
and can define usual EE, $S'_{\rm A}, S'_{\rm B},$ and $S'_{\rm A\cup B}$.
The mutual information is given by $I_{A;B} = S'_{\rm A} + S'_{\rm B} - S'_{\rm A\cup B}$. 

\begin{figure}[Htb]
 \includegraphics[clip, width=1\columnwidth, height=0.55\columnwidth]{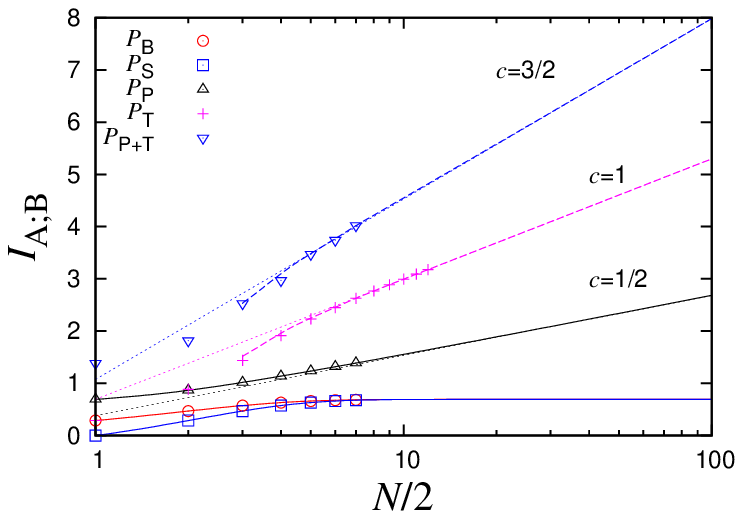} 
    \caption{(Color online) Mutual information $I_{A;B}$ in the symmetry projection operators for 1D systems. 
Plotted symbols are numerical results by using the singular value decomposition. 
The solid (dashed) curves mean the analytic (asymptotic) results with the cubic spline interpolation.
The dotted lines are fitted by $c\log(N)+$const. in the large $N$ region.
\comment{IM: 二次元やるべき}
}
\label{matual_ee}
\end{figure}

It is easy to calculate $I_{A;B}=0$ for the identity operator $\hat{1}$, and the inversion operator $\hat{T}_{\rm B}$.  Then, this can be a suitable measure for an entropy generation of the projections.
We denote, for example, the mutual information $I_{A;B}$ of a projection $\hat{P}_{\rm S}$ as $I^{\rm S}$.
After straight forward calculation,
one can calculate
analytic results of $I_{A;B}$,\cite{note_I}
which are depicted by solid lines in Fig. \ref{matual_ee}. 
For $I^{\rm T}$, the asymptotic results in the thermodynamic ($N\rightarrow \infty$) limit are depicted as a dashed line in Fig. \ref{matual_ee}.
In addition, $I^{\rm P+T}$ in large $N$ region gets close to the sum of $I^{P}+I^{T}$, which is quite reasonable.
The system size $N=16$ has been done is almost in the asymptotic region.
We focus on the mutual information $I_{A;B}$ in the thermodynamic limit.
The mutual information $I^{\rm S}$ and $I^{\rm B}$ in the thermodynamic limit is $\log 2$
as seen in Fig. \ref{matual_ee}.
On the other hand, $I^{\rm P}$, $I^{\rm T}$, and $I^{\rm P+T}$ are increasing logarithmically with respect to $N$ in large $N$ region as $I_{A;B}\sim c\log(N)$+const. 
The coefficients $c$ are $c=1/2$ for $\hat{P}_{\rm P}$, $c=1$ for $\hat{P}_{\rm T}$, and $c=3/2$ for $\hat{P}_{\rm P+T}$.
This implies that these three projection operators generate significant EE in any 1D chain, because the EE of $S_{\rm A}$ is increasing logarithmically at most in the 1D system even if the system is critical. 
Moreover, as a result of the mutual information, $\hat{P}_{\rm T}$ generates larger entanglement than $\hat{P}_{\rm P}$.

In summary, we investigate the effect of the symmetry projection operators on the MPS, TPS, TTN, and MERA in the spin $S=1/2$ Heisenberg 1D chain and 2D square lattice of 16 sites. 
Calculating the EE and the variation energy in the network states, we obtain the conclusion that the variational energy is refined monotonically by applying each symmetry projection operator which reflects the global symmetry of the exact ground state. 
This behavior appears irrespective of the structure of the network and will be valid for larger systems. 
Especially, the translational invariant projection operator generates EE most effectively in the both 1D and 2D systems. 
This result is consistent with the analytic result of maximum of the EE and the mutual information generated by the projection operators,
which means that the property of the projection itself is important.
This analysis gives the reason why larger EE in the variational calculation is obtained by the translational symmetric projection than by the other projections.
In addition, as a numerical result, dimensionality can be absorbed in the translational symmetric projection.

In this sense, there is a possibility of more effective projections for higher dimensions.
Application to the orthogonal dimer model~\cite{JPCM.15.327} with frustration is one of future issues, because this model having an exact ground state can be defined in any higher dimensions theoretically and an excited state or ground state in another phase is interesting. 
Even in the case of 1D systems, the uniform MPS applied to the translational symmetry braking state and local spin twisting projections are analyzed recently~\cite{JPSJ.80.023001}, 
which will give another possibility of symmetry projections.
Another future problem is to apply this scheme of the global-symmetry projection operators to larger systems, 
because straightforward optimization procedure used in this paper cannot deal the system size more than that of numerical exact diagonalization. 
It is also interesting to consider magnetic-field dependence of optimal tensor network states, because the full ferromagnetic state has no EE.
%

This work was supported in part by a Grant-in-Aid for JSPS Fellows and Grant-in-Aid No. 20740214, Global COE Program (Core Research and Engineering of Advanced Materials-Interdisciplinary Education Center for Materials Science) from the Ministry of Education, Culture, Sports, Science and Technology of Japan. 


\begin{thebibliography}{10}
\bibitem{note_index_correspondence}
In 2D calculation, the 1D tensor networks are reconfigured on the following 1D
  system. The correspondence between the 2D spin index $\sigma_{i j}$ on the $4
  \times 4$ square lattice and 1D index $\sigma_{k}$ is as follows:
  $\sigma_{14} = \sigma_{1,1}$, $\sigma_{15} = \sigma_{2,1}$, $\sigma_2 =
  \sigma_{3,1}$, $\sigma_3 = \sigma_{4,1}$, $\sigma_{13} = \sigma_{1,2}$,
  $\sigma_{16} = \sigma_{2,2}$, $\sigma_1 = \sigma_{3,2}$, $\sigma_4 =
  \sigma_{4,2}$, $\sigma_{12} = \sigma_{1,3}$, $\sigma_{9} = \sigma_{2,3}$,
  $\sigma_{8} = \sigma_{3,3}$, $\sigma_5 = \sigma_{4,3}$, $\sigma_{11} =
  \sigma_{1,4}$, $\sigma_{10} = \sigma_{2,4}$, $\sigma_{7} = \sigma_{3,4}$, and
  $\sigma_6 = \sigma_{4,4}$.
\bibitem{note_DMRG}
The eneregies $E$ are $-7.134817$ ($-7.138986$, $-7.139465$, $-7.139943$) for
  the DMRG of $\max \chi=15$ (16, 17, 18), and $-7.136659$ ($-7.139653$) for
  the MERA+T+P (MERA+T+P+B+S) of $\bm{\chi}=(3,2,1)$.
\bibitem{note_II}
When the DPS is $\ket{\Psi} = ( \sum_{ {\bm \sigma}_{\rm A} } f( s^z_{\rm A} )
  \ket{ {\bm \sigma}_{\rm A} }) ( \sum_{{\bm \sigma}_{\rm B}} f( s^z_{\rm B} )
  \ket{ {\bm \sigma}_{\rm B} })$ $\ket{\tilde{\Psi}}$ of $\hat{P}_{\rm P}$ has
  $\log(N/2 + 1)$, where $s^z_{\rm A} = \sum_{i \in A} \sigma_i$, $s^z_{\rm B}
  = \sum_{i \in B} \sigma_i$ and $f(s^z_{\rm A}) \propto 1/\sqrt{
  _{N_A}C_{N_A/2+s^z_{\rm A}} }$.
\bibitem{note_I}
Analytic results are $I_{\rm A;B}=-\sum_{i=0}^{N/2} f_i \log f_i$ for
  $\hat{P}_{\rm P}$, $I_{\rm A;B}=\log2 - \log(1 + 2^{-N/2})$ for $\hat{P}_{\rm
  B}$, and $I_{\rm A;B}=\log2 - \log(1 + 2^{-N/2 - 1})$ for $\hat{P}_{\rm S}$,
  where $f_i = (N/2)!^4 /[ N! (N/2 - i)!^2 i!^2]$.
\end{thebibliography}

\end{document}